\documentclass[
reprint,
twocolumn,
english,
aps,
prb,
superscriptaddress,
bibnotes,
amsmath,
amssymb,
floatfix]{revtex4-1}

\usepackage[utf8]{inputenc}
\usepackage[T1]{fontenc}
\usepackage{graphicx}
\usepackage{dcolumn}
\usepackage{siunitx}

\usepackage[colorlinks=true, allcolors=blue]{hyperref}
\usepackage[mathlines]{lineno}
\usepackage{hyperref}
\usepackage{marvosym}

\graphicspath{{figures/}}

\begin{document}

\title{Integrated tunable green light source on silicon nitride}
%\title{Feedback-enabled turnkey mode-locked and Q-switched operation}

\author{Gang Wang}
\affiliation{École Polytechnique Fédérale de Lausanne, Photonic Systems Laboratory (PHOSL), Lausanne, Switzerland}
\author{Ozan Yakar}
\affiliation{École Polytechnique Fédérale de Lausanne, Photonic Systems Laboratory (PHOSL), Lausanne, Switzerland}
\author{Xinru Ji}
\affiliation{École Polytechnique Fédérale de Lausanne, Laboratory of Photonics and Quantum Measurements (LPQM), Lausanne, Switzerland}
\author{Marco Clementi}
\affiliation{École Polytechnique Fédérale de Lausanne, Photonic Systems Laboratory (PHOSL), Lausanne, Switzerland}
\affiliation{Dipartimento di Fisica “A. Volta”, Università di Pavia, Via A. Bassi 6, 27100 Pavia, Italy}
\author{Ji Zhou}
\affiliation{École Polytechnique Fédérale de Lausanne, Photonic Systems Laboratory (PHOSL), Lausanne, Switzerland}
\author{Christian Lafforgue}
\affiliation{École Polytechnique Fédérale de Lausanne, Photonic Systems Laboratory (PHOSL), Lausanne, Switzerland}
\author{Jiaye Wu}
\affiliation{École Polytechnique Fédérale de Lausanne, Photonic Systems Laboratory (PHOSL), Lausanne, Switzerland}
\author{Jianqi Hu}
\affiliation{École Polytechnique Fédérale de Lausanne, Laboratory of Photonics and Quantum Measurements (LPQM), Lausanne, Switzerland}
\author{Tobias J. Kippenberg}
\affiliation{École Polytechnique Fédérale de Lausanne, Laboratory of Photonics and Quantum Measurements (LPQM), Lausanne, Switzerland}
\author{Camille-Sophie Brès}
\email{camille.bres@epfl.ch}
\affiliation{École Polytechnique Fédérale de Lausanne, Photonic Systems Laboratory (PHOSL), Lausanne, Switzerland}

\date{\today}

\begin{abstract}
\noindent
Integrated green light sources are essential for telecommunications and quantum applications, while the performance of current on-chip green light generation is still limited in power and tunability. 
In this work, we demonstrate green light generation in silicon nitride microresonators using photo-induced second-order nonlinearities, achieving up to 3.5 mW green power via second-harmonic generation and densely tunable over a 29 nm range. 
In addition, we report milliwatt-level all-optical poling (AOP) threshold, allowing for amplifier-free continuous-wave AOP. 
Furthermore, we demonstrate non-cascaded sum-frequency generation, leveraging the combination of AOP and simultaneous coherent frequency combs generation at 1 \si{\micro\meter}.
Such comb-assisted AOP enables switching of the green light generation over an 11 nm range while maintaining the pump within a single resonance.
The combination of such highly efficient photo-induced nonlinearity and multi-wavelength AOP enables the realization of low-threshold, high-power, widely-tunable on-chip green sources.

\end{abstract}

\maketitle

\section*{Introduction}

\noindent 
Green light sources, typically defined by emission wavelengths within the 510--560 nm range, have wide applications in fields such as quantum photonics \cite{bernien2013heralded,wu2019programmable,clark2021high}, material processing \cite{haubold2018laser}, and underwater communication \cite{sticklus2018optical}. 
While semiconductor lasers are widely available in the blue and red wavelength regions, efficient green laser sources are difficult to produce and are often limited in tunability. As such, nonlinear conversion is the standard approach to generate green light, which has been widely adopted in tabletop systems based on crystals.

Recently, progress in photonic integration has led to the realization of compact and scalable on-chip green light sources based on second-order ($\chi^{(2)}$) and third-order ($\chi^{(3)}$) nonlinearities, leveraging frequency conversion processes such as third-harmonic generation (THG) \cite{corcoran2009green,vijayakumar2024phase,levy2011harmonic,wang2016frequency,Jung2014GreenredIR,surya2018efficient}, optical parametric oscillation (OPO) \cite{lu2020chip,domeneguetti2021parametric,sun2024advancing}, sum-frequency generation (SFG) \cite{ling2022third,yakar2022generalized,hu2022photo}, and second-harmonic generation (SHG) \cite{chen2025continuous,porcel2017photo}.
On-chip green light via THG has been generated in silicon photonic crystal waveguides \cite{corcoran2009green}, silicon nitride (Si\textsubscript{3}N\textsubscript{4}) waveguides and microrings \cite{vijayakumar2024phase,levy2011harmonic,wang2016frequency}, AlN microrings \cite{Jung2014GreenredIR}, and composite SiN/AlN microrings \cite{surya2018efficient} where up to 49 \si{\micro\watt} at 514 nm was achieved. 
$\chi^{(3)}$ OPO-based green sources in Si\textsubscript{3}N\textsubscript{4} microrings pumped near 750 nm \cite{lu2020chip,domeneguetti2021parametric,sun2024advancing,Lu2024Emergingintegratedlaser} have shown significantly improved tunability over the visible spectrum, achieving a maximum on-chip green power of tens of \si{\micro\watt} \cite{domeneguetti2021parametric}.

Frequency conversion via $\chi^{(2)}$ nonlinearity is generally a more efficient approach. Cascaded SHG and SFG with a telecom pump has been exploited in periodically poled thin-film lithium niobate microring \cite{ling2022third} reaching up to 334 \si{\micro\watt} at 520 mn, yet the efficient generation has been restricted to a single resonance. SHG in peridiodically poled lithium tantalate waveguide \cite{chen2025continuous} resulted in 1.87 mW at 532 nm, with a 0.40 nm generation bandwidth. 
In amorphous materials, despite lacking intrinsic $\chi^{(2)}$ nonlinearity, 
an effective $\chi^{(2)}$ can be induced through all-optical poling (AOP), firstly observed in silica fibers \cite{anderson1991model,dianov1994photoinduced,osterberg1986dye,margulis1995imaging} then in Si\textsubscript{3}N\textsubscript{4} waveguides and microresonators \cite{billat2017large,porcel2017photo,hickstein2019self,yakar2022generalized,lu2021efficient,nitiss2022optically,hu2022photo,zhou2024self,li2025down}.
Based on multi-photon absorption interference involving three photons $\omega_{1,2,3}$, a directional photocurrent $j_{\rm ph}$ is generated via the coherent photogalvanic effect (PGE), promoting the build-up of an electrostatic field $E_{\rm DC}$. 
The steady state of $E_{\rm DC}$ can be expressed as $E_{\rm DC}=-j_{\rm ph}/\sigma$, where $j_{\rm ph}$ and photoconductivity $\sigma$ are functions of the electric fields oscillating at $\omega_{1,2,3}$ frequencies\cite{yakar2022generalized}. 
The static electric field induced by the AOP process exhibits a periodic distribution along the propagation axis with a wavevector $\Delta k=k_3-k_1-k_2=2\pi/\Lambda$, where $k_{1,2,3}$ represent the wavevectors of $\omega_{1,2,3}$ and $\Lambda$ is the spatial period of $E_{\rm DC}$. 
As such, an equivalent periodically modulated second-order nonlinearity $\chi^{(2)}_{\rm eff}=3\chi^{(3)}E_{\rm DC}$ is inscribed, satisfying the first-order QPM condition for the $\chi^{(2)}$ frequency conversion $\omega_3=\omega_1+\omega_2$. 
While most studies have been carried out in the 1550 nm telecom band generating a near-infrared second-harmonic (SH), green light generation from a 1 \si{\micro\m} pulsed pump in an optically poled Si\textsubscript{3}N\textsubscript{4} waveguide has also been demonstrated \cite{porcel2017photo}. In addition, cascaded SFG after photo-induced SHG from a telecom pump have also been shown in Si\textsubscript{3}N\textsubscript{4} waveguides and microresonators\cite{yakar2022generalized,hu2022photo}. 
So far, the green light generation in 
Si\textsubscript{3}N\textsubscript{4} has been limited to a sub-mW level \cite{lu2020chip,domeneguetti2021parametric}, and the trade-off between power and tunability is still challenging to solve.

\begin{figure*}[htb!]
    \centering
    \includegraphics[width=1 \textwidth]{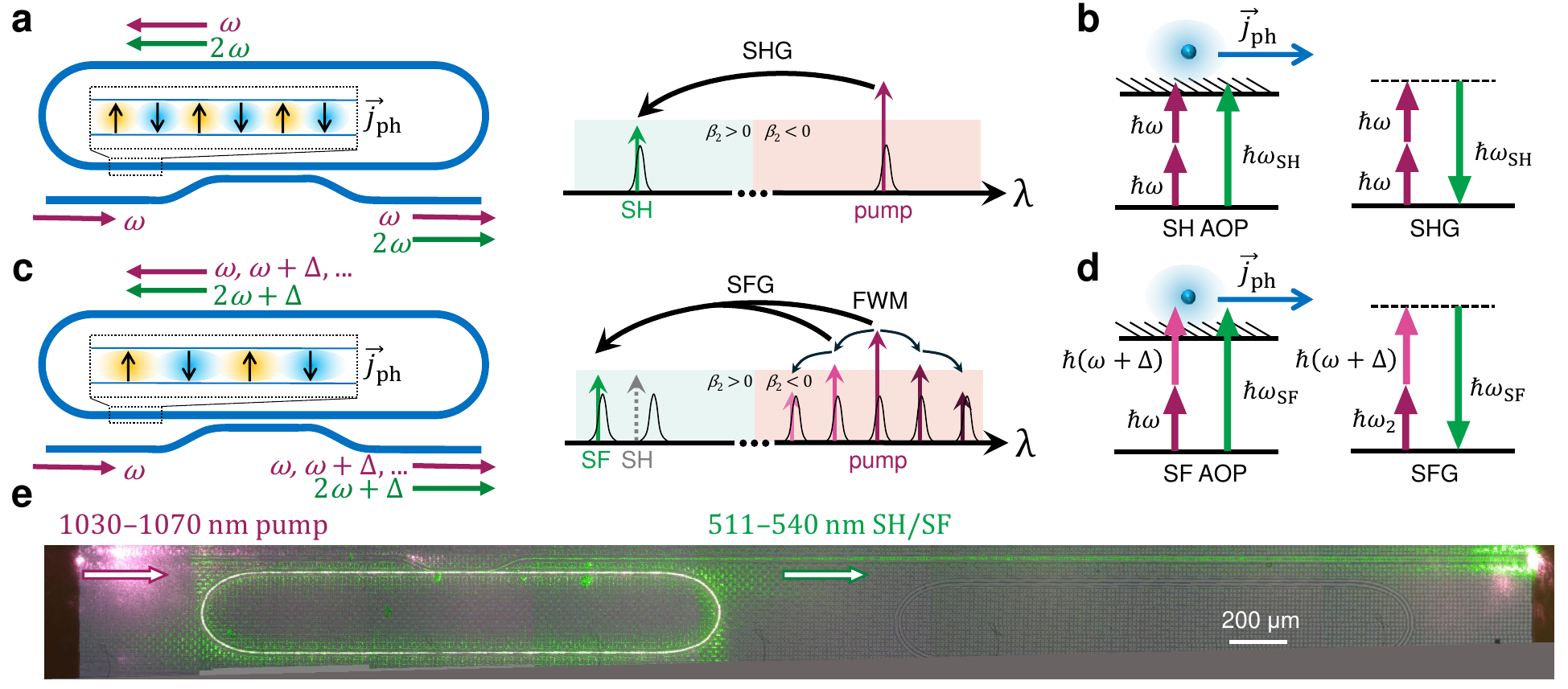}
    \caption{
        \textbf{All-optical-poling-enabled green light generation in Si\textsubscript{3}N\textsubscript{4} microresonators} 
        \textbf{a} Schematic of SHG in the Si\textsubscript{3}N\textsubscript{4} microresonator when doubly resonant condition is met. The dashed box shows the zoomed-in view of the optically induced photocurrent distribution. Black Lorentzian curves in the right figure represent resonances of the microresonator.
        \textbf{b} Illustration of SH-AOP-induced photocurrent generation and energy conservation of SHG. The FH and SH waves collectively contribute to the generation of a directed photocurrent through multi-photon absorption interference, thereby inscribing the built-in periodic electric field which further interacts with FH to generate SH.
        \textbf{c} Schematic of SFG in the Si\textsubscript{3}N\textsubscript{4} microresonator when triply resonant condition is selectively met among two comb lines and the SF signal. At pump frequency the waveguide exhibits anomalous dispersion, allowing the injected pump at $\omega$ frequency to be converted into multiple comb lines with a frequency difference of $\Delta$ through cascaded four-wave mixing. SF AOP can generate its own $\chi^{(2)}$ grating different from the SH gratings, as shown in the dashed box.
        \textbf{d} Illustration of SF-AOP-induced photocurrent generation and the energy conservation of SFG.
        \textbf{e} Microscope image of the device under operation. Green light is generated in the ring and out-coupled to the right. With 1030--1070 nm pump, green light with a wavelength of 511--540 nm can be densely achieved.
    }
    \label{fig:fig1}
\end{figure*}

%Our results
In this work, we demonstrate an integrated green light source based on AOP in Si\textsubscript{3}N\textsubscript{4} microresonators, achieving up to 3.5-mW green power, densely tunable over a 29 nm range. 
Furthermore, we report the AOP threshold as low as 4.5 mW of pump power required to trigger continuous-wave (CW) AOP process, which bypasses the need for optical amplifiers. 
By leveraging the anomalous dispersion of our microrings at 1 \si{\micro\meter}, we also investigate the interactions of photo-induced $\chi^{(2)}$ and $\chi^{(3)}$-based frequency comb generation.
Moreover, we show that comb-assisted AOP allows for non-cascaded SFG, which is widely switchable over a range of 11 nm by slight pump detuning in a single pumped resonance. 
These results advance on-chip efficient green light sources that are low-threshold, high-power and widely tunable.

\section*{Results}

\begin{figure*}[hbt!]
    \centering
    \includegraphics[width=1 \textwidth]{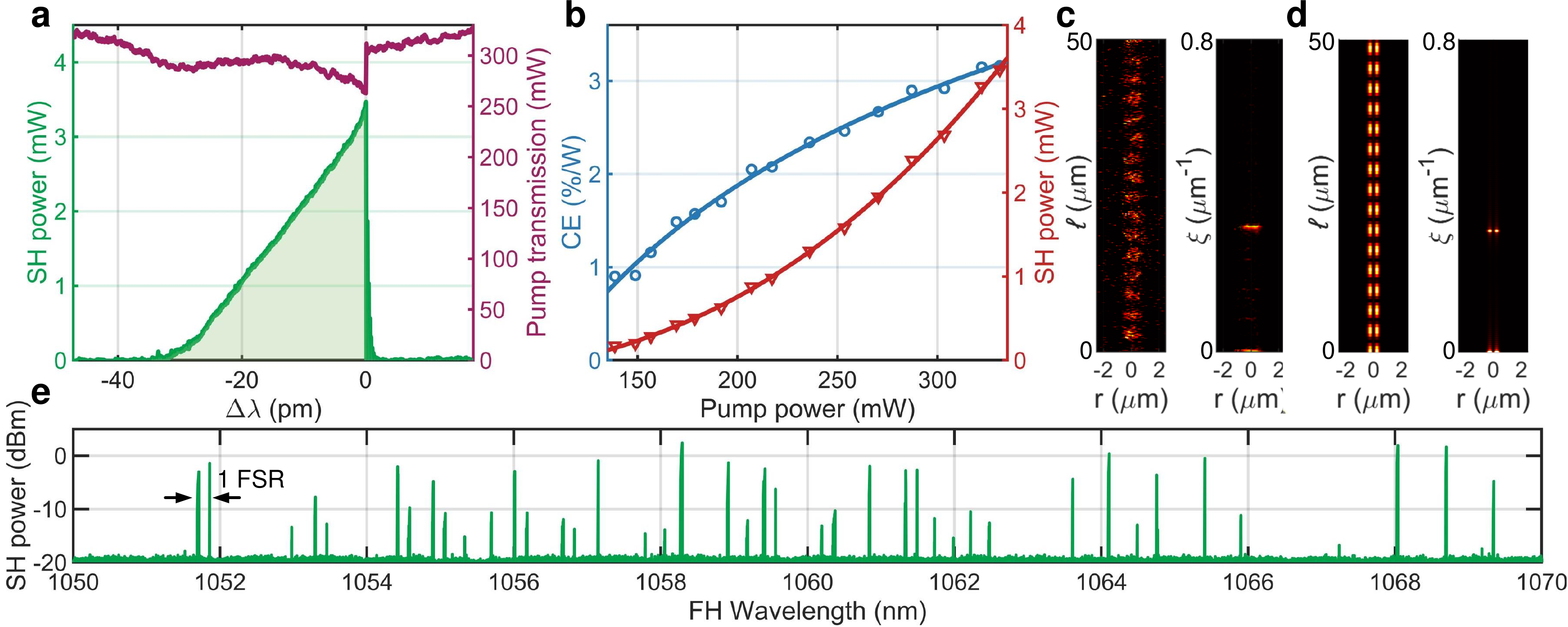}
    \vspace{-5mm}
    \caption{
    \textbf{Reconfigurable high-power green light generation.}
    \textbf{a} Optimized nonlinear resonance sweep. The green and purple curves represent the on-chip SH power and the pump transmission.
    \textbf{b} SH conversion efficiency and output power as a function of pump power. 
    \textbf{c} Experimental two-photon microscope image and extracted spatial distribution of the optimized AOP situation, which are in good agreement with \textbf{d} simulations involving the interaction between the fundamental FH mode and SH2.
    \textbf{e} Broadband SHG via reconfigurable AOP. Approximately 40\% of resonances can achieve AOP. FSR: free spectral range.
    }
    \label{fig:fig2}
\end{figure*}

\noindent 
The mechanism for green light generation in our microresonator is illustrated in Fig.~\ref{fig:fig1}, where the coherent photogalvanic effect and four-wave mixing (FWM) play key roles.
The Si\textsubscript{3}N\textsubscript{4} racetrack microresonator in our study has a waveguide cross-section of 1.3$\times $0.9~$\si{\micro\meter}^2$. In the pump wavelength (fundamental harmonic - FH) range around 1 \si{\micro\m}, the resonator exhibits anomalous dispersion for the fundamental mode. At the SH wavelength range, the fundamental mode and all higher-order modes identified in our poling process exhibit normal dispersion. Details on the device layout and dispersion calculations are given in Supplementary Note 1 and shown in Fig.~S1.

Figures \ref{fig:fig1}a and \ref{fig:fig1}b describe the photo-induced SHG process initiated by AOP. A tunable CW 1-$\si{\micro\meter}$ pump (1030--1070 nm) is injected into the bus waveguide and couples to the resonator through a directional coupler as the fundamental TE mode. 
When the FH and SH are both in resonance, i.e., doubly resonant, SHG can be achieved through the following positive feedback in Fig.~\ref{fig:fig1}b, 
until an equilibrium is achieved between the photogalvanic current and the photoconductivity-induced drift current \cite{yakar2022generalized,zhou2024self}. The photocurrent $\vec{j}_{\rm ph}$ and the electric field $\vec{E}_{\rm DC}$ have a spatial distribution as shown in the zoomed view of Fig.~\ref{fig:fig1}a. 
Given that the waveguide is multimode at SH band, AOP can occur via the interaction of the fundamental pump mode and SH higher-order transverse modes, resulting in an adaptation of the grating period as well as its geometrical shape \cite{nitiss2022optically}.

Figures~\ref{fig:fig1}c and \ref{fig:fig1}d depict the interactions FWM and photo-induced SFG processes. At the pump wavelength, anomalous dispersion can facilitate the generation of optical frequency combs via the intrinsic $\chi^{(3)}$ nonlinearity, giving rise to multiple comb lines with a frequency interval of $\Delta$. When a triply resonant condition among two comb lines and the SF signal is met, for example between \si{\omega}, \si{\omega}+$\Delta$ and 2\si{\omega}+$\Delta$, this non-cascaded SFG can be realized similar to the aforementioned SHG process. QPM grating inscription and SFG can be enhanced mutually through a positive feedback as shown in Fig.~\ref{fig:fig1}d. This comb-mediated AOP leads to self-sustained SFG, which indicates the SFG writes its own grating instead of relying on any pre-inscribed gratings. This process effectively addresses the frequency limitations encountered in both degenerate and cascaded SFG methods, resulting in improved tunability.
Figure~\ref{fig:fig1}e shows a typical photograph of the device under operation, where the green light is efficiently generated in the cavity and coupled out to the bus waveguide.

\subsection*{High-power and reconfigurable on-chip green source}

\noindent
First, AOP for SHG is carried out in a Si\textsubscript{3}N\textsubscript{4} racetrack microresonator with a free spectral range (FSR) of 50 GHz and a 494 $\si{\nano\meter}$ gap between the ring and the bus. 
Detailed characterization, experimental setup, and mode profiles of participating SH modes (defined as SH1 to SH4) are provided in the Supplementary Note 1. 
The loaded quality factor (Q) of a typical pump resonance is measured to be 0.8$\times$$10^6$ as shown in Fig.~S2, with intrinsic Q of 12.8$\times$$10^6$, overcoupled at the pump.

Figure~\ref{fig:fig2}a presents the optimized fast resonance sweep with a speed of 3.7 nm/s at 1068 nm under a pump power of 25.2 dBm in the bus waveguide, where the on-chip SH power reaches a maximum of 5.4 dBm (3.5 mW).
The pump transmission exhibits an atypical thermal triangle shape, attributed to the Fano lineshape of the resonance \cite{limonov2017fano}. 
The SH power increases monotonically with detuned wavelength, showing a different behavior compared to the typical case observed at 1.55 $\si{\micro\meter}$ \cite{nitiss2022optically}. 
One possible explanation is that the effective detuning of SH resonance $\delta'$ (difference between SH resonance frequency and twice that of the pump frequency) is changing from negative to zero during the resonance sweep, thereby ensuring the AOP condition ($\delta'<0$) \cite{zhou2024self} from the beginning of the process. 
This observation suggests that the thermal shift of the FH resonance is faster than that of the SH resonance. %, different from earlier works \cite{nitiss2022optically, zhou2024self}. 
The conversion efficiency (CE = $P_{\rm SH}/P_{\rm pump}^2$) and the on-chip SH power as a function of the on-chip pump power are shown in Fig.~\ref{fig:fig2}b, reaching 3.2 \mbox{\%/W} and 3.5 mW, respectively.

The inscribed grating can stay inside the Si\textsubscript{3}N\textsubscript{4} microring after AOP \cite{hickstein2019self}, which can be measured by two-photon microscropy (TPM) \cite{nitiss2022optically,hickstein2019self}.
For the resonance under test, the TPM pattern reveals the spatial distribution and associated spatial frequency shown in Fig.~\ref{fig:fig2}c, which aligns simulation considering the interaction between the transverse mode pair FH-SH2 ($\xi=0.28\si{\micro\meter}^{-1}$), as shown in \ref{fig:fig2}d. 
Other TPM patterns in different resonances, written by other transverse mode pairs FH and SH modes from SH1 to SH4 are also observed (Supplementary Note 3).

AOP occurs for a multitude of resonances when the pump wavelength is swept from 1050 and 1070 nm.
Figure~\ref{fig:fig2}e illustrates the reconfigurable SHG under a stabilized temperature of 51$^\circ$C. The pump power is 24.8 dBm in the bus waveguide and the wavelength of the pump laser is swept with a speed of 0.01 nm/s.
Among all 105 pumped resonances, 42 exhibited detectable SHG.
A detailed wavelength scan conducted within the 1060–1070 nm range reveals that approximately 63\% of resonances yield green light output, as illustrated in Fig.~S4.

\subsection*{Low-AOP-threshold on-chip green source}

\noindent The AOP process is known to exhibit a poling threshold \cite{zhou2024self}, which at 1550 nm telecom wavelength has been observed to decrease with finesse. 
A low threshold is highly desirable for fully integrated devices, as this would eliminate the need for amplifier-assisted poling. 
Although there have been several reports of on-chip self-injection locking SHG in AOP devices operating at C/L band \cite{clementi2023chip,li2023high}, they required pre-writing of gratings with an amplified pump to reach the necessary poling threshold. 
This limitation restricts some applications and adds to operational complexity, hindering the ability to independently reconfigure or maintain effective $\chi^{(2)}$ gratings. 
As high finesse lowers the AOP threshold, with record of few mW was reported in a 1-THz high-Q Si\textsubscript{3}N\textsubscript{4} microring \cite{lu2021efficient,lu2021considering} operating at 1550 nm. 
However, large FSR rings significantly restrict the tunability of the process while the threshold significantly increases in larger microresonators at the telecom wavelength.

\begin{figure}[ht!]
    \centering
    \includegraphics[width=0.47 \textwidth]{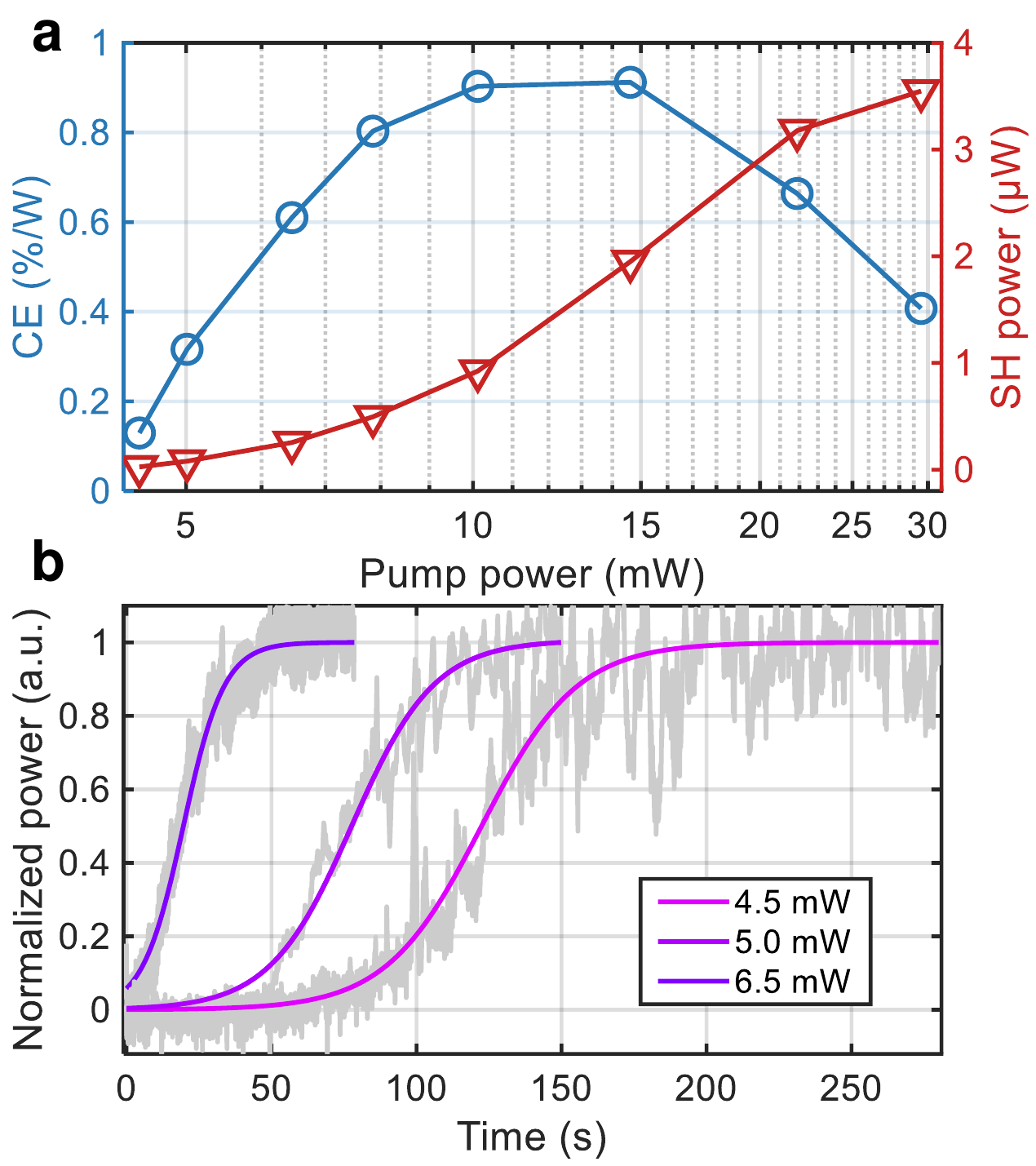}
    \vspace{-5mm}
    \caption{
    \textbf{Low-threshold AOP with milliwatt-level pump.}
    \textbf{a} Conversion efficiency and output power of SH as a function of pump power. The threshold for AOP and harmonic generation is measured at 4.5 mW.
    \textbf{b} Growth of generated SH power over time under different pump power. The growing SH power proves the capability of self-assisted low-threshold AOP.
    }
    \label{fig:fig3}
\end{figure}

\begin{figure*}[ht!]
    \centering
    \includegraphics[width=1\textwidth]{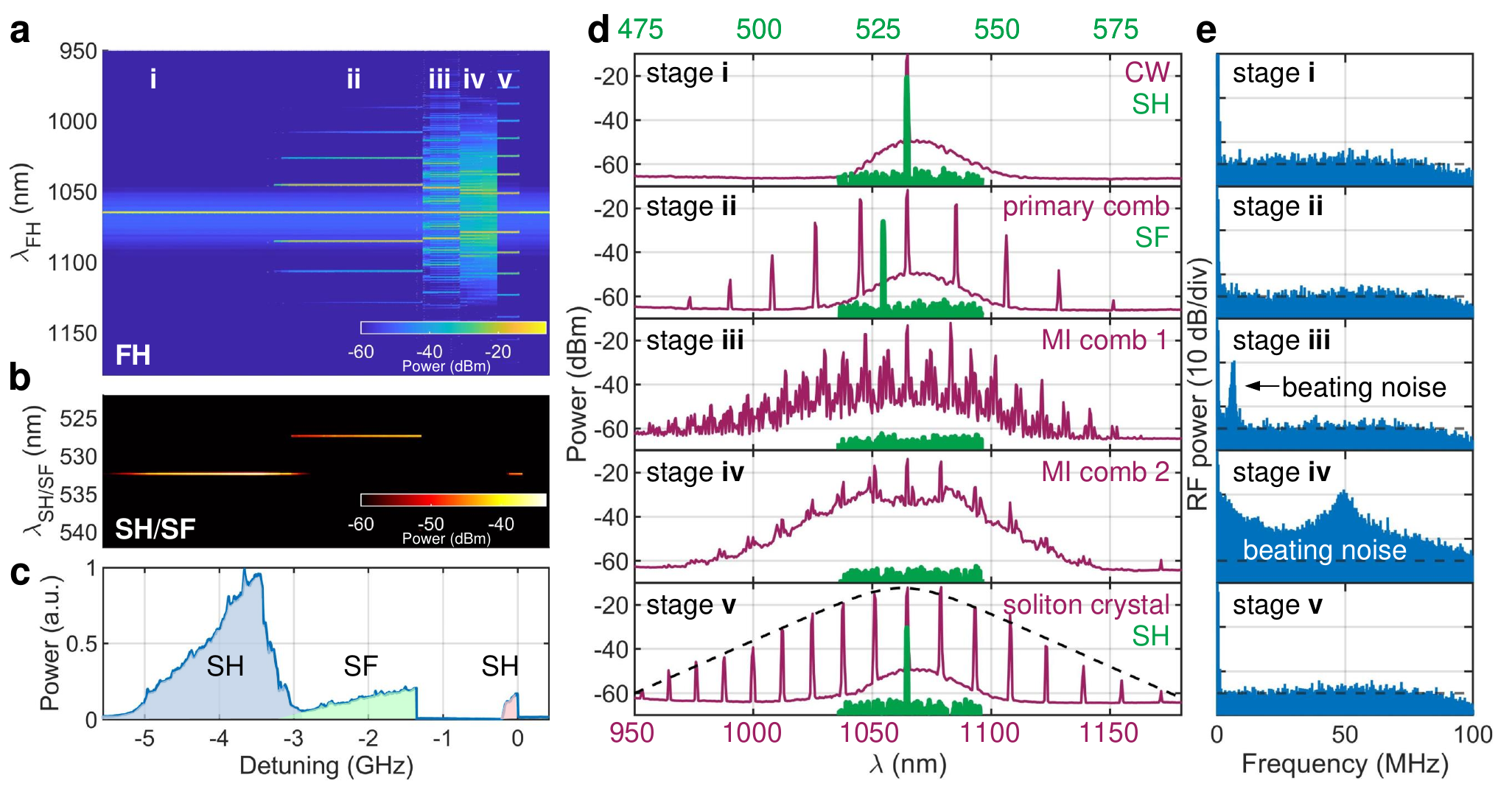}
    \caption{
    \textbf{Kerr-comb assisted all-optical poling and erasing.}
    \textbf{a-c} Evolution of FH spectra, green spectra, and green output power in a single linear resonance sweep. FH experiences five stages, three of which are coherent and facilitate AOP while the other two lead to grating erasure. The corresponding spectra of FH and SH/SF in five stages are shown and marked in \textbf{d}.
    \textbf{e} Measurement of beating noise in a 100 MHz span. The noise in stage \textbf{iii} and \textbf{iv} shows the beating signal of incoherent MI combs.
    }
    \label{fig:fig4}
\end{figure*}

However, the physical principles of AOP and of the coherent PGE, suggest that operating the system at shorter pump wavelengths offers a significant opportunity for lowering the AOP threshold. The coherent PGE relies on charge excitation, where shorter wavelength light is more prone to be absorbed and excite the charges. Therefore, the photogalvanic coefficient at 1 \si{\micro\meter} should be higher than that at telecom wavelength, enabling low-threshold AOP even in small FSR rings.
This property could bypass the need for an amplifier and facilitate fully on-chip reconfigurable AOP. 
We investigated the onset of AOP in our 50 GHz ring with a gap of 567 nm exhibiting a loaded Q of 1.8$\times$$10^6$ (see detailed characterization in Supplementary Note 1). The results are shown in Fig.~\ref{fig:fig3}a, plotting the measured variations in the CE (blue) and the SH power (red) as a function of the pump power in the bus waveguide. 
Despite a significantly lower finesse (320) than that of the previous demonstration \cite{lu2021efficient} (6.2$\times$$10^3$), the AOP threshold in this work is still lower and is estimated at 4.5 mW in the bus waveguide. The stronger photogalvanic effect at short wavelength contributes to the lowering of the poling threshold.
Beyond that value, the CE increases steadily until the pump power reaches approximately 15 mW, after which it starts to decrease. This decrease in CE indicates that, with the increased SH powers generated, the inscribed DC field is reduced due to the more rapid increase in photoconductivity compared to photocurrent.

To confirm that low-power green light generation is indeed self-assisted rather than the reading of a pre-written gratings, the measured growth of SH power as a function of time at low pump powers (4.5, 5, and 6.5 mW) is illustrated in Fig.~\ref{fig:fig3}b, normalized to the maximum power obtained at each case. 
The gray and colored curves show the measured and sigmoid-fit SH power. 
At a pump power of 5 mW, the half-rise time of SH power from the sigmoid fit is approximately 78 seconds. 
When the power in the bus waveguide increases to 6.5 mW, the poling time decreases to 18 seconds. 
At 10 mW, the half-rise time is less than 1 second, with considerable CE. 
A 10 mW pump power represents a readily achievable threshold for on-chip lasers, demonstrating the capability of this microresonator device to facilitate fully integrated on-chip AOP.

\subsection*{Frequency-comb-mediated all-optical poling}

\noindent In a microresonator with anomalous dispersion, Kerr combs such as primary combs, MI combs, and different types of solitons states can be realized. 
The primary combs and solitons are coherent, namely they exhibit a specific phase relation between comb lines, while the noise-driven MI combs are incoherent \cite{erkintalo2014coherence,mitchell1996self}. 
Coherent combs can write and sustain QPM gratings \cite{hickstein2019self,porcel2017photo,hu2022photo}, while incoherent combs, due to their high power, high phase noise, and complex modes involved, could contribute to the erasure of gratings. Here, we demonstrate the poling and erasure of the gratings in a Si\textsubscript{3}N\textsubscript{4} microring based on coherent and incoherent Kerr combs. 

During a single resonance sweep from blue to red detuning, we measure the optical spectrum in both the FH and SH band, as well as the radio frequency (RF) signals from the 100-MHz near-infrared photodetector at the output of our device. 
Fig.~\ref{fig:fig4}a shows the evolution of the attenuated FH spectra recorded by the optical spectrum analyzer (OSA), where five stages appear, including CW, primary comb, first and second MI combs, and soliton crystal. 
Corresponding to the five stages of FH, green signals are observed and marked in Figs.~\ref{fig:fig4}b (OSA spectra) and \ref{fig:fig4}c (green light power). Figs~\ref{fig:fig4}d and \ref{fig:fig4}e display the detailed overlayed dual-axis FH and SH spectra, as well as beating noise of the FH signal of these five stages. 

In stage \textbf{i}, the FH is a 1064.6-nm CW as shown in Fig.~\ref{fig:fig4}d. A green signal at 532.3 nm gradually appears, corresponding to SHG, indicating that a doubly resonant condition is met. We can see in Fig.~\ref{fig:fig4}d that the SH signal is precisely located at half the wavelength of the FH. Fig.~\ref{fig:fig4}e shows the noise floor of FH detected by a 100-MHz photodetector.

With the FH wavelength approaching the effective resonance wavelength (detuning approaching zero), the coupled FH power increases and modulation instability gain is triggered, enabling primary comb generation indicated in stage \textbf{ii}. 
Correspondingly, the SH signal gradually diminishes during the primary comb generation process.
As the SH at 532.3 nm decreases, a SF signal at 527.3 nm gradually replaces the SH, indicating that a triply resonant condition for SFG is met. 
In the early stage of primary comb generation, SH and SF exist concurrently for a short range until SH completely disappears, suggesting that the SH grating is gradually erased while the SF grating is written. This indicates that SFG experiences a more favorable AOP condition in comparison to SHG during the detuning change, leading to a lower AOP threshold and a competitive advantage. 
The overlayed dual-axis spectra in stage \textbf{ii} show that the SF signal is located between the FH and the first-order sideband at shorter wavelength (1044.9 nm). 
The noise measurement in Fig.~\ref{fig:fig4}e shows that the primary comb remains in a low-noise coherent state.

\begin{figure*}[ht!]
    \centering
    \includegraphics[width=1\textwidth]{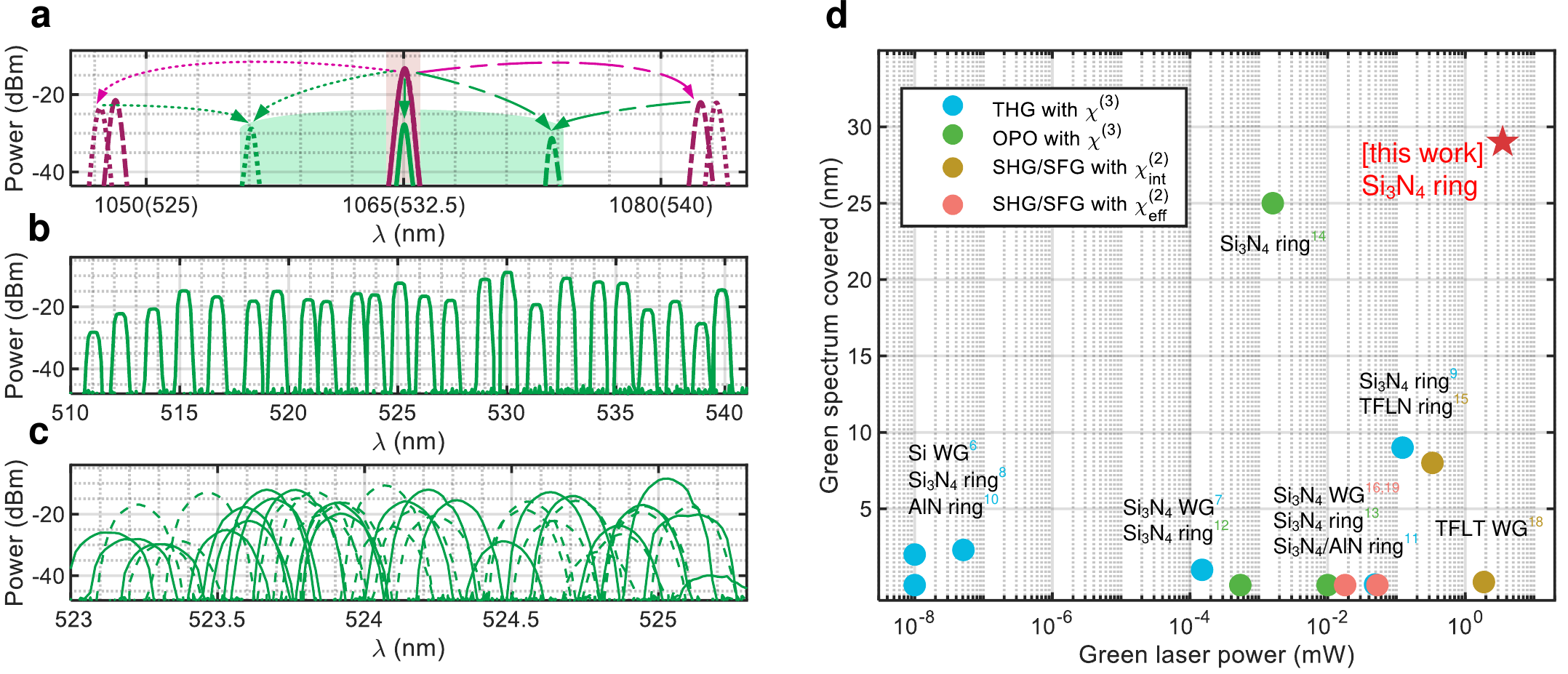}
    \caption{
    \textbf{Performance of the green light source.}
    \textbf{a} Overlapped spectra of the 11-nm-switchable green light generation and the associated 1-\si{\micro\m} pump. The solid, dahsed, and dash-dotted traces represent three different states recorded by the OSA when detuning is changed within a single pump resonance.
    \textbf{b} Tunable range of 29 nm from 511 to 540 nm at a fixed temperature of 50$^\circ$C. 
    \textbf{c} Green signals obtained in a 2-nm range at 50$^\circ$C (solid line) and 40$^\circ$C (dashed line). 
    \textbf{d} Enhanced output power and spectral coverage compared to other integrated frequency-conversion green sources. WG: waveguide; TFLN: thin-film lithium niobate; TFLT: thin-film lithium tantalate.
    }
    \label{fig:fig5}
\end{figure*}

With detuning further approaching zero, the FH enters stages \textbf{iii} and \textbf{iv}, corresponding to MI comb states, where no upconverted signals can be observed. 
Fig.~\ref{fig:fig4}d shows that the FH spectrum exhibits chaotic states, and the corresponding noise spectrum displays significant beating signals between different sets of incoherently related combs. The absence of any upconverted signals can be due to unfavorable resonant condition at the SH band. It can also be ascribed to a disproportionate rise in photocurrent and photoconductivity within the high-power and high-phase-noise MI state, resulting in the pump power falling below the AOP threshold. Note that under different condition, and in the less conductive telecom range, 1550-nm MI combs can be upconverted under close to group-velocity matched condition between the FH and SH bands\cite{clementi2024ultrabroadband}.

The FH finally enters stage \textbf{v}, where a soliton crystal is observed, along with an upconverted signal, probably triggered by another low threshold doubly resonant condition being met. The overlayed spectrum in Fig.~\ref{fig:fig4}d shows that the FH comb exhibits a sech$^2$-shaped envelope, in Fig.~\ref{fig:fig4}e that the FH goes back to a coherent low-noise state.

The combination of coherent comb generation and AOP as shown in Fig.~\ref{fig:fig4} serves as the first direct experimental confirmation of self-assisted non-cascaded SFG. 
Achieving AOP in Si\textsubscript{3}N\textsubscript{4} waveguides only requires the involving coherent lights exceeding the poling threshold, while in microresonators, a triply (or doubly for the case of degenerate processes) resonant condition is also necessary. 
Experimentally, both degenerate SFG (SHG), and cascaded SFG (effective THG) have been demonstrated \cite{hu2022photo}. 
In these cases, the AOP comes from the interaction of a pump and its direct harmonics. 
However, self-assisted non-cascaded SFG results from SF-AOP triggered by two coherent signals not related through harmonics.

In practice, such non-cascaded SF-AOP is constrained primarily by AOP competition. The presence of the two pumps can impede the SF AOP process if each of them can pole its own SH $\chi^{(2)}$ gratings. 
Notably, when the two beams are not identically strong, which is often the case, the stronger beam tends to dominate the competition resulting in regular SHG instead of SFG. 

In our device, the non-cascaded coherent pump can be realized through Kerr combs generation by leveraging the anomalous dispersion at the pump, while the AOP threshold is significantly reduced to the mW level while maintaining small FSR. 
Given the presence of multiple comb lines with significant powers, the SF product can be switched depending on the achievable triply resonant condition.
Fig.~\ref{fig:fig5}a shows a typical case for switchable green light generation with a pump near 1065 nm. As the pump is slightly tuned, the green signals can be switched in a range of 11 nm as the primary comb state varies along with detuning conditions. The solid, dashed, and dash-dotted traces represent the overlapped FH and green spectra recorded under three detuning conditions. Detailed spectra evolution with pump detuning can be found in Fig.~S7. 

When pump wavelength is varied in the range of 1030--1070 nm at a fixed temperature of 50$^\circ$C, the tunable range of CW green signals spans from 511 to 540 nm as shown in Fig.~\ref{fig:fig5}b, covering half of the green spectrum. The green signal in Fig.~\ref{fig:fig5}b is selectively sampled to ensure that the interval is no more than 2 nm, but in fact, the green signal can be much denser. To demonstrate the density of the green signals, Fig.~\ref{fig:fig5}c shows all reconfigurable green signals obtained in a 2-nm span by varying pump wavelength under fixed temperature, where the signal density is 10.5 nm$^{-1}$ at 50$^\circ$C and 11.6 nm$^{-1}$ at 40$^\circ$C with the help of the reconfiguration, which is more than 6 times denser than OPO-based tunable green source 1.6 nm$^{-1}$ \cite{sun2024advancing}.

\section*{Discussion}
\noindent
Figure~\ref{fig:fig5}d summarizes the performance of on-chip green laser sources in terms of power and tunability. 
Compared to previous studies, both the 3.5-mW green laser power and the 29-nm tunable range achieved in this work show the leading performance. 
The high spectrum density of generated green light also allows for flexible frequency tuning in our device.

Based on our theoretical modeling of AOP in resonant systems, the AOP threshold is inversely proportional to the microresonator finesse \cite{zhou2024self}. 
Despite reaching a mW-level AOP threshold, there is still substantial potential to reduce the AOP power by over an order of magnitude. Our microresonator, with a 50 GHz FSR and non-critical coupling, suggests that a microring with a 1-THz FSR could achieve an AOP threshold below 1 mW. 
Such ultralow-threshold showcases the potential of this approach for developing a fully integrated self-injection-locked green laser source.

Additionally, by combining AOP with intrinsic $\chi^{(3)}$ processes, frequency-comb mediated non-cascaded SFG offers additional capability for wavelength tuning of the green light signal within a fixed pump resonance, opening up new possibilities for applications such as self-reference-stabilized comb.

In summary, we have developed an integrated green source through photo-induced second-order nonlinearities, achieving record performance in terms of power and tunability. 
This work significantly enhances the capabilities of Si\textsubscript{3}N\textsubscript{4} for visible light generation. 
From a physical standpoint, we provide the experimental evidence linking optical frequency comb to AOP conditions, demonstrate the feasibility of non-cascaded SF AOP for the first time, and utilize this approach to access a broader green light spectrum.

\vspace{0.5cm}
\noindent\textbf{Acknowledgements}
\medskip
\begin{footnotesize}

\noindent This work is supported by Swiss National Science Foundation (SNSF grant MINT 214889).
\end{footnotesize}

\vspace{0.5cm}
\noindent\textbf{Competing interests}
\medskip
\begin{footnotesize}

\noindent The authors declare no conflict of interest.

\end{footnotesize}

\vspace{0.5cm}
\noindent\textbf{Data availability}
\medskip
\begin{footnotesize}

\noindent The data and code that support the plots within this paper and other findings of this study are available from the corresponding authors upon reasonable request.

\end{footnotesize}

\medskip
\begin{footnotesize}

\end{footnotesize}

\bibliographystyle{apsrev4-1}
\bibliography{references}

\end{document}

\title{Supplementary Information for: \\Integrated tunable green light source on silicon nitride}

\author{Gang Wang}
\affiliation{École Polytechnique Fédérale de Lausanne, Photonic Systems Laboratory (PHOSL), Lausanne, Switzerland}
\author{Ozan Yakar}
\affiliation{École Polytechnique Fédérale de Lausanne, Photonic Systems Laboratory (PHOSL), Lausanne, Switzerland}
\author{Xinru Ji}
\affiliation{École Polytechnique Fédérale de Lausanne, Laboratory of Photonics and Quantum Measurements (LPQM), Lausanne, Switzerland}
\author{Marco Clementi}
\affiliation{École Polytechnique Fédérale de Lausanne, Photonic Systems Laboratory (PHOSL), Lausanne, Switzerland}
\affiliation{Dipartimento di Fisica “A. Volta”, Università di Pavia, Via A. Bassi 6, 27100 Pavia, Italy}
\author{Ji Zhou}
\affiliation{École Polytechnique Fédérale de Lausanne, Photonic Systems Laboratory (PHOSL), Lausanne, Switzerland}
\author{Christian Lafforgue}
\affiliation{École Polytechnique Fédérale de Lausanne, Photonic Systems Laboratory (PHOSL), Lausanne, Switzerland}
\author{Jiaye Wu}
\affiliation{École Polytechnique Fédérale de Lausanne, Photonic Systems Laboratory (PHOSL), Lausanne, Switzerland}
\author{Jianqi Hu}
\affiliation{École Polytechnique Fédérale de Lausanne, Laboratory of Photonics and Quantum Measurements (LPQM), Lausanne, Switzerland}
\author{Tobias J. Kippenberg}
\affiliation{École Polytechnique Fédérale de Lausanne, Laboratory of Photonics and Quantum Measurements (LPQM), Lausanne, Switzerland}
\author{Camille-Sophie Brès}
\email{camille.bres@epfl.ch}
\affiliation{École Polytechnique Fédérale de Lausanne, Photonic Systems Laboratory (PHOSL), Lausanne, Switzerland}

%\date{\today}

\maketitle

\section{Si\textsubscript{3}N\textsubscript{4} microresonator devices}
\begin{figure*}[h!]
    \centering
    \includegraphics[width=1\textwidth]{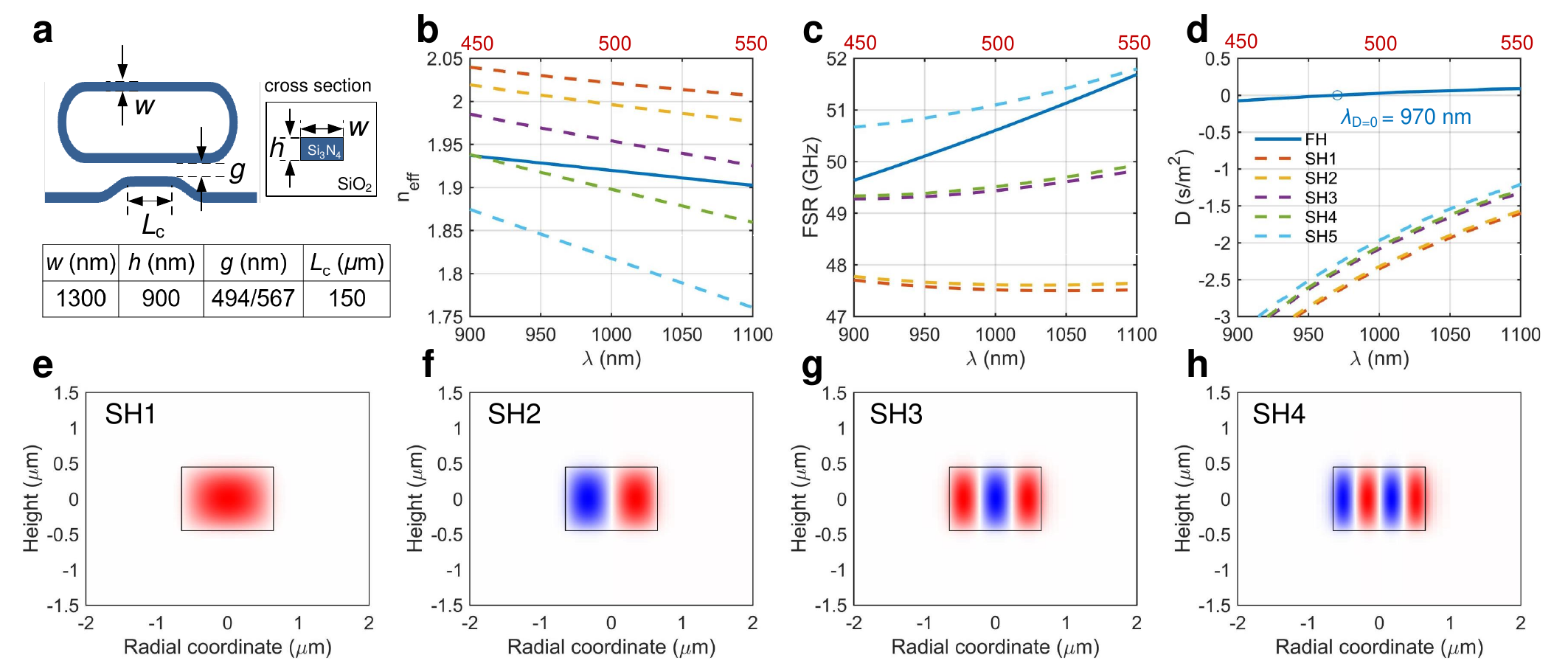}
    \caption{
    \textbf{Design and simulation of the Si\textsubscript{3}N\textsubscript{4} microresonators.} 
    \textbf{a}
    Schematic of the Si\textsubscript{3}N\textsubscript{4} racetrack microresonator. The racetrack microresonator is coupled to the bus waveguide with a directional coupler. The parameters are listed on the table. $w$: width; $h$: height; $g$: gap; $L_{\rm c}$: coupler length.
    \textbf{b-d} 
    Simulated effective refractive index, FSR, and dispersion parameter $D$ of fundamental mode at FH wavelength  and first five SH (SH1-SH5) modes.
    \textbf{e-h}
    Simulated TE-polarized mode amplitude distributions of SH1-SH4.
    }
    \label{fig:figS1}
\end{figure*}

\noindent
The Si\textsubscript{3}N\textsubscript{4} microresonators used in this study have a racetrack geometry, coupled to a bus waveguide via a directional coupler, with  design parameters shown in Fig. \ref{fig:figS1}a. 
Two different coupler gaps 494 and 567 nm are used with other parameters remaining the same. Both the microresonator waveguide and the bus waveguide have cross-sectional dimensions of 1.3$\times $0.9~$\si{\micro\meter}^2$. 
In this geometry, the simulated effective index ($n_\mathrm{eff}$), free spectral range (FSR), and dispersion parameter ($D$) of the transverse electric (TE) modes at fundamental harmonic (FH) and second harmonic (SH) are illustrated in Figs. \ref{fig:figS1}b-\ref{fig:figS1}d. 
Notably, the fundamental TE mode exhibits anomalous dispersion ($D > 0$) at wavelengths longer than 970 nm, enabling comb generation through third-order nonlinearity. 
The TE transverse mode profiles of FH and SH1-3 are depicted in Figs. \ref{fig:figS1}e-\ref{fig:figS1}h. 
These mode profiles and their effective refractive indices are subsequently used to simulate the theoretical pattern of the optically inscribed gratings.

\begin{figure*}[h!]
    \centering
    \includegraphics[width=1\textwidth]{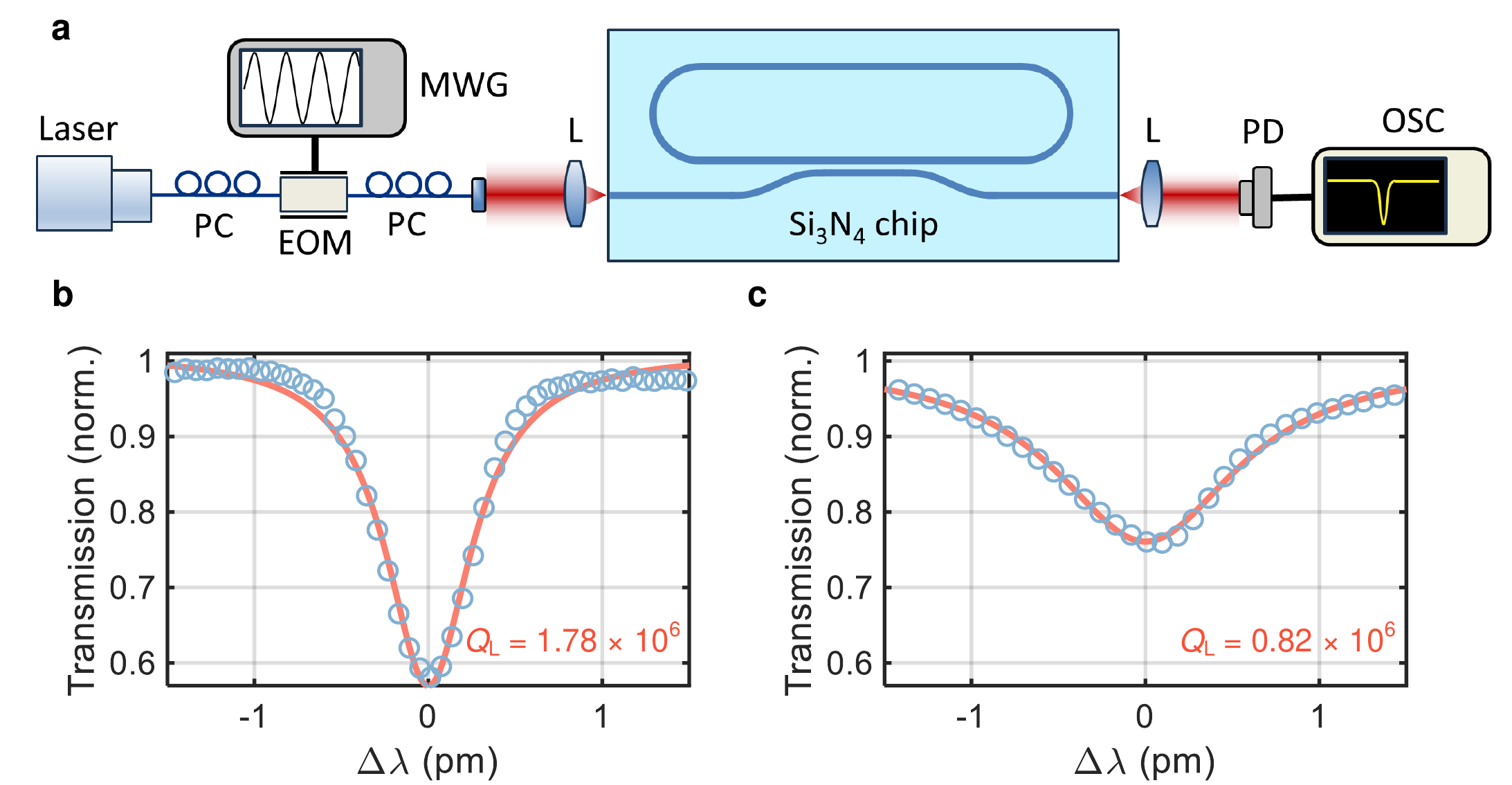}
    \caption{
    \textbf{Setup and characterization of the Si\textsubscript{3}N\textsubscript{4} microresonators.} 
    \textbf{a} Experimental setup for characterizing the resonance linewidth of microresonators. PC: polarization controller; EOM: electro-optic modulator; MWG: microwave generator; C: collimator; L: lens; PD: photodetector; OSC: oscilloscope. Resonance linewidth measurement of \textbf{b} the 567-nm and \textbf{c} 494-nm gap microresonator devices. The measured linewidths (blue points) are fitted with a Lorentzian function (red curve), showing the loaded Q factor ($Q_{\rm L}$) of $1.78\times10^6$ and $0.82\times10^6$, respectively.
    }
    \label{fig:figS2}
\end{figure*}

Figure \ref{fig:figS2}a shows the setup for characterizing resonance linewidth using the sideband modulation technique \cite{li2012sideband}. The laser output is polarization-maintaining and tunable within 1030--1070 nm. The first polarization controller is added to set the polarization state to TM, which is required for phase modulation in the electro-optic modulator (EOM). The microwave generator gives a 2-GHz signal to the EOM, which creates two sidebands as reference to calibrate the resonance sweep. The laser is then set to TE polarization and coupled into the bus waveguide through a free-space collimator and a lens. The output laser is collected by a lens and sent to a photodetector connected to the oscilloscope.

Figures \ref{fig:figS2}b and c show the resonance linewidth at 1065-nm of two resonators with 567-nm and 494-nm gap, respectively. The loaded Q factors $Q_{\rm L}$ of the two resonators are $1.78\times10^6$ and $0.82\times10^6$, as well as the minimum transmittance $T_{\rm min}$ are 0.56 and 0.76 respectively. To calculate the intrinsic and coupling Q factor $Q_{\rm i}$ and $Q_{\rm c}$, we use the equation:

\begin{equation}
Q_{\rm i},Q_{\rm c}=\frac{2Q_{\rm L}}{1\pm\sqrt{T_{\rm min}}},
\label{eq1}
\end{equation}

The $\pm$ in the denominator of the formula depends on the coupling condition to the microresonator: if over-coupled $Q_{\rm i}>Q_{\rm c}$, and if under-coupled $Q_{\rm i}<Q_{\rm c}$. For the microresonator with 567-nm gap, we have $Q_{\rm i1},Q_{\rm c1}=2.03\times10^6,14.27\times10^6$. Similarly, we have $Q_{\rm i2},Q_{\rm c2}=0.87\times10^6,12.77\times10^6$ for the second microresonator with 494-nm gap. Based on the Lumerical FTDT simulation of the directional coupler, the coupling Q factor $Q_{\rm c}$ of two cases should be approximately $2.53\times10^6$ and $0.98\times10^6$, respectively. Comparing with the measured $Q_{\rm i},Q_{\rm c}$,we can tell the two microresonators are both over-coupled at 1 $\si{\micro\meter}$ region, meaning the coupling losses exceed the intrinsic losses of the microresonators. However, for SH fundamental transverse mode, the directional coupler only couples less than 0.01\% of green light, together with the large intrinsic loss, resulting in significantly under-coupled condition. Given the better FH extinction ratio of 567-nm-gap ring and better SH extraction ability of 494-nm-gap ring, the two resonators are respectively used to investigate Kerr-comb-related all-optical poling (AOP) and efficient green laser generation.

\section{Experimental setup}
\begin{figure*}[h!]
    \centering
    \includegraphics[width=1\textwidth]{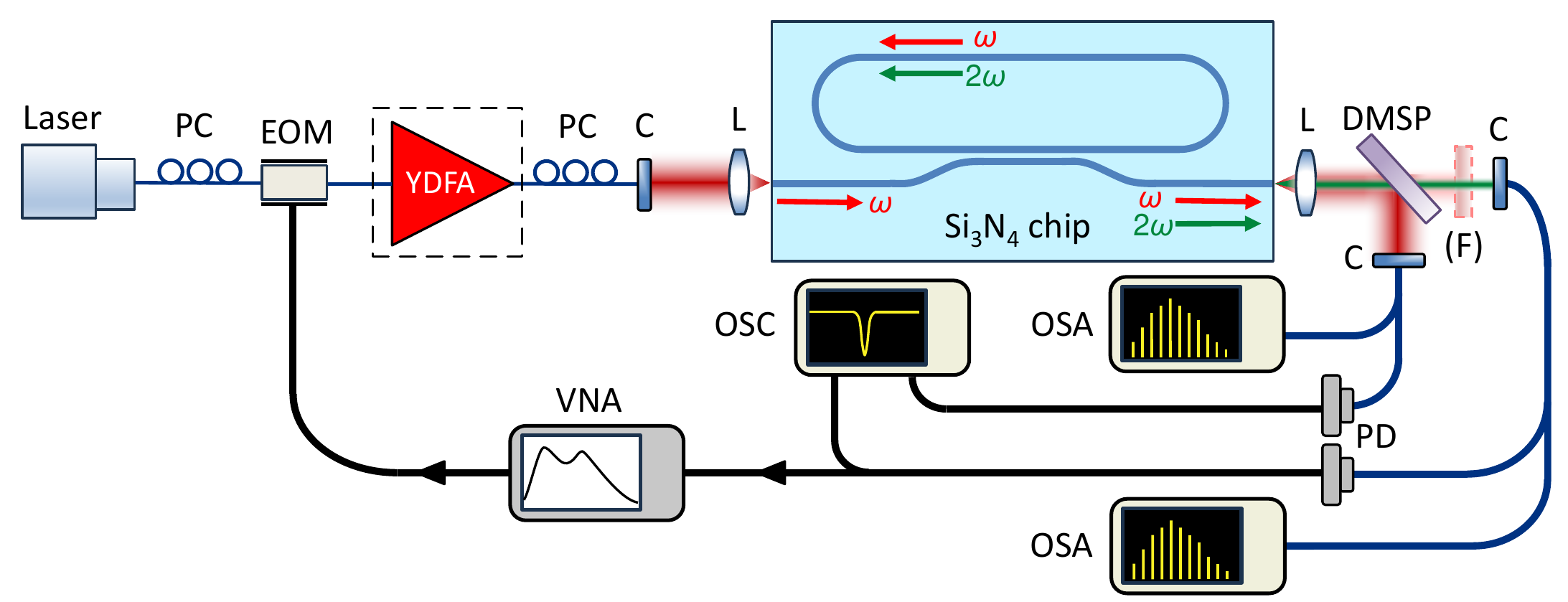}
    \caption{
    \textbf{Experimental setup for all-optical poling, measuring, and probing the SH response.} 
    An ytterbium-doped fiber amplifier (YDFA) is added after the EOM. The dashed box represents that the YDFA is not used in some of the experiments. A short-pass dichroic mirror (DMSP) is added after the outcoupling lens to separate FH and SH signal. The FH signal leaks a small proportion through the DMSP and can be filtered out completely using several low-pass filters (F). Two collimators collect the output FH and SH separately and send them to the oscilloscope (OSC) and OSAs. A vector network analyzer (VNA) is used to probe the FH and SH resonance through sending a weak modulation signal (50-1500 MHz) to the EOM and detecting the RF response of the SH.
    }
    \label{fig:figS3}
\end{figure*}

\noindent
The setup for AOP and detection is shown in Fig. \ref{fig:figS3}. Compared to the linewidth measurement setup in Fig. \ref{fig:figS3}a, the EOM here is much weakly driven by the VNA that sends 50--1500 MHz signal. The VNA then receives the response of SH to measure the transfer function. An ytterbium-doped fiber amplifier (YDFA) is inserted between two polarization controllers.  The FH laser (red arrow) is coupled to the ring and interacts with the defect-induced weak SH to generate periodic photocurrent in the waveguide to pole an intracavity grating, which in turn enhance the SH through positive feedback. The generated SH is then coupled out of the ring through the directional coupler and emitted from the facet of the chip. A shortpass dichroic mirror is used to separate FH from SH, and additional filters in the SH path can be used to remove any residual FH. Both FH and SH are collected by two collimators and sent to different photodetectors and instruments to measure their power, spectra, and transfer function. The resolution of infrared and visible optical spectrum analyzers (OSAs) are set to 0.5 and 0.1 nm, respectively. The taper loss is measured to be 3.3 dB by comparing the measured Fabry–Pérot interference with the theoretical case without any taper loss. The loss of the C-coating lens, dichroic mirror, and filters is measured to be 3.1 dB in total for SH.

\section{Fine scan of reconfigurable AOP and $\chi^{(2)}$ measurement}
\begin{figure*}[h!]
    \centering
    \includegraphics[width=1\textwidth]{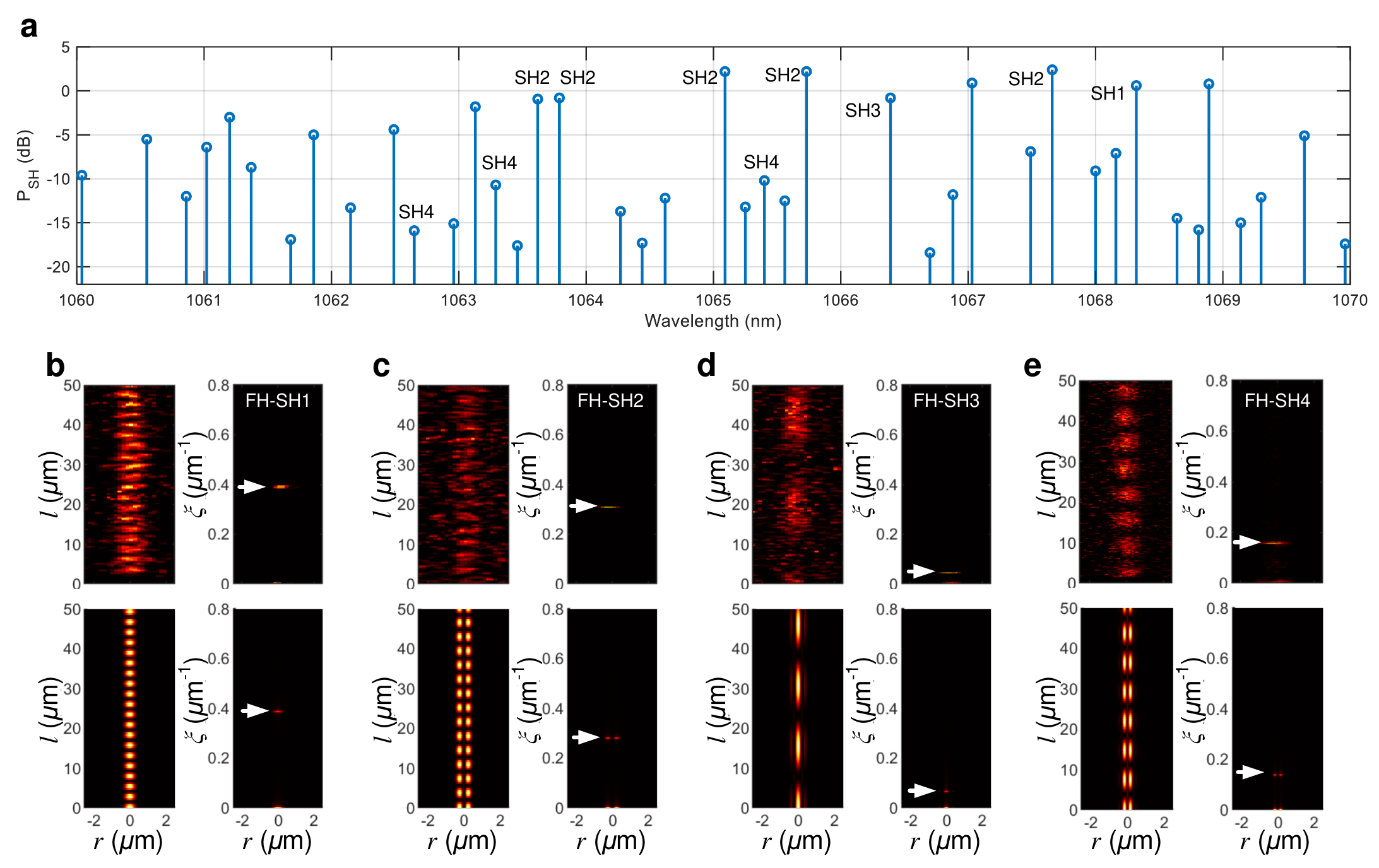}
    \caption{
    \textbf{TE polarization resonance scan with reconnfigurable intracavity gratings.} 
    \textbf{a} Generated SH power during the wide scan of pump from 1060 to 1070 nm, where 63\% of the resonances can realize AOP. 
    \textbf{b-e} Measured TPM gratings distribution (top left) and their spatial period (top right), as well as the simulation counterpart (bottom).
    }
    \label{fig:figS4}
\end{figure*}

\noindent To show the broadband SHG and the identification of mode pair interactions of the 494-nm-gap device, we carry a finer TE polarized scan in the range of 1060--1070 nm region for fixed  pump power and temperature same as Fig.~2e. Figure \ref{fig:figS4}a shows the on-chip SH power during pump wavelength scan. Within all the 62 fundamental FH resonances, 39 of them gives reconfigurable SH response. We selected several resonance of interest to do two-photon microscopy, where the typical TPM patterns observed (FH--SH1 to FH--SH4) are shown in Fig. \ref{fig:figS4}b--\ref{fig:figS4}e.

\begin{figure*}[h!]
    \centering
    \includegraphics[width=1\textwidth]{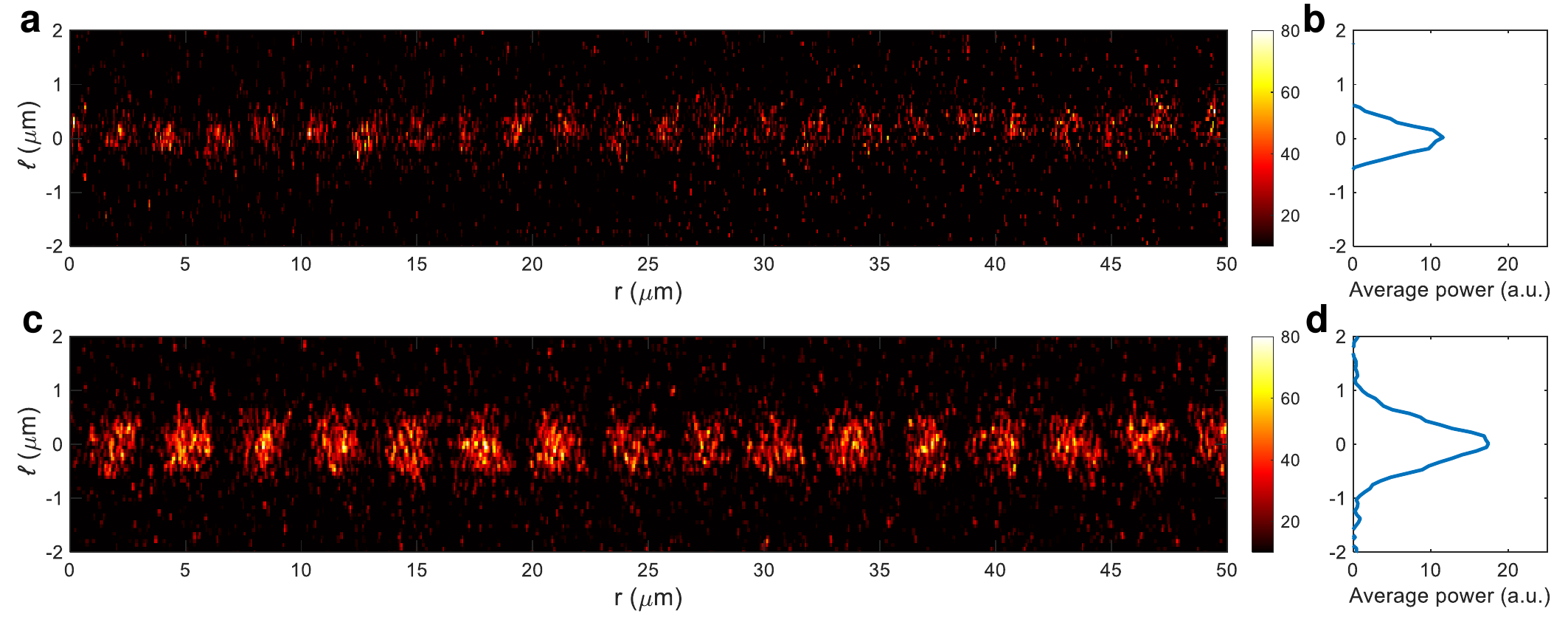}
    \caption{
    \textbf{Intracavity $\chi^{(2)}_{\rm eff}$ measurement.} 
    \textbf{a} and \textbf{c} TPM response of a poled waveguide and the poled microresonator.
    \textbf{b} and \textbf{d} show the averaged TPM response of \textbf{a} and \textbf{c} over distance.
    }
    \label{fig:figS5}
\end{figure*}

We estimate the intracavity $\chi^{(2)}_{\rm eff}$ by comparing the TPM response of a poled ring to that of a waveguide under the same TPM parameters including laser power, wavelength, focus, etc. The intensity of TPM response is proportional to the square of $\chi^{(2)}_{\rm eff}$.
Figures \ref{fig:figS5}a and \ref{fig:figS5}b show the TPM pattern of the poled waveguide and the averaged response over distance in the focused area, respectively. Similarly, the maximum TPM response of the poled ring is shown in Figs. \ref{fig:figS5}c and \ref{fig:figS5}d. The response of the resonator is 1.51 times that of the waveguide, which means that $\chi^{(2)}_{\rm eff}$ of the ring is 1.23 times $\chi^{(2)}_{\rm eff}$ in the waveguide. The $\chi^{(2)}_{\rm eff}$ of the poled waveguide is estimated to be 0.024 pm/V from its conversion efficiency \cite{yakar2022generalized}. Therefore the intracavity $\chi^{(2)}_{\rm eff}$ is approximately 0.03 pm/V.

\section{Additional proofs of sum-frequency all-optical poling}
\begin{figure*}[h!]
    \centering
    \includegraphics[width=0.9\textwidth]{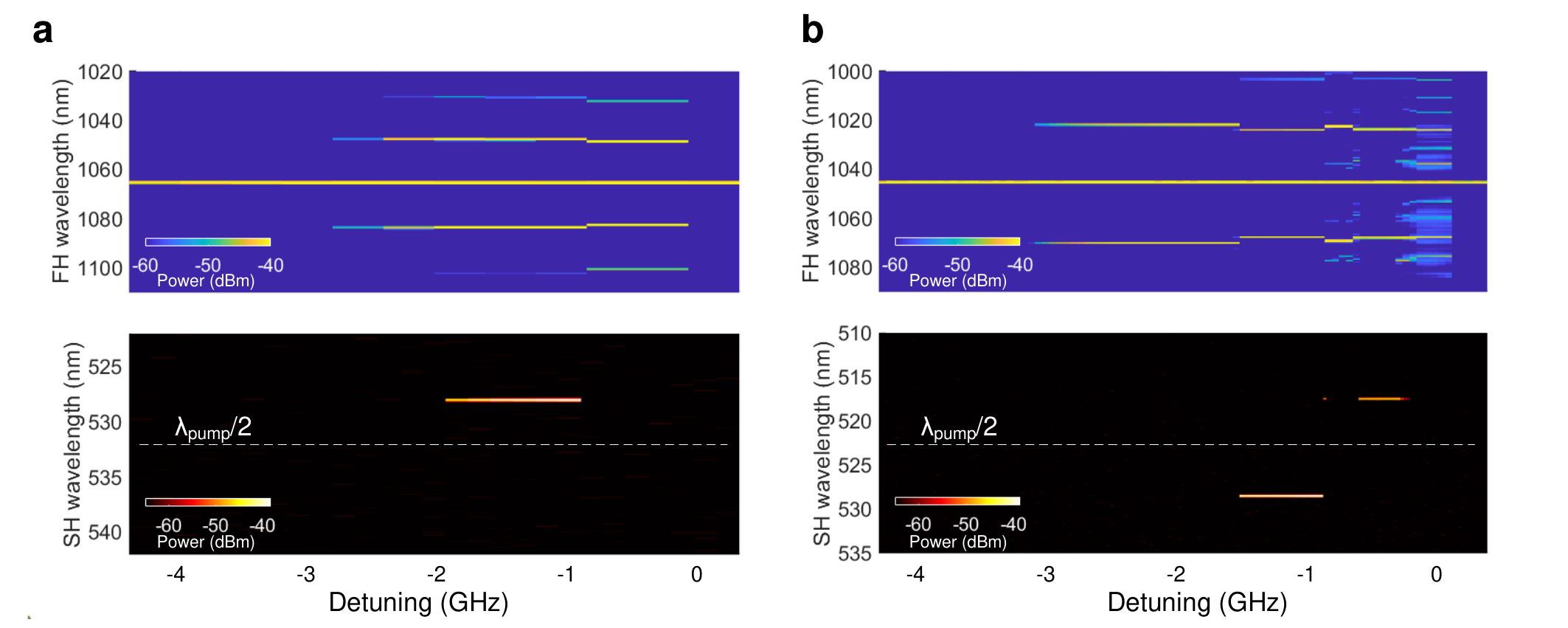}
    \vspace{-5 mm}
    \caption{
    \textbf{Sum-frequency AOP without SHG.} 
    \textbf{a and b} Evolution of FH (top) and SF (bottom) spectra in single linear resonance sweep at 1065 and 1045 nm, respectively. When doubly resonant condition between SF and SH is not met, the triply resonant condition between FH, primary comb sideband, and SF can be realized, which enables the SF AOP. These results further demonstrate that SH and SF generation depend on distinct grating structures, making them incapable of supporting each other's generation in our device.
    }
\vspace{-5 mm}
    \label{fig:figS6}
\end{figure*}

\noindent
Figure \ref{fig:figS6} shows the observation of sum-frequency generation (SFG) during a slow resonance scan at 1065 nm and 1045 nm respectively. The top patterns show the evolution of FH from CW to a series of primary combs based on detuning conditions. The bottom patterns show the evolution of SH/SF spectrum, with the dashed white line marking the half wavelength of the FH pump. Any possible gratings before the measurement were bleached.

After the generation of certain primary combs, SF signals are observed approximately 5 nm away from half of FH wavelength. The wavelengths of the SF signal match well with a quarter of the sum of the pump and a combline. Notably, previous studies have reported SFG driven by SHG gratings, but the SFG observed here originates from self-induced AOP gratings supported by several reasons: (i) No SHG signal can be detected during the resonance sweep, ruling out the existence of SH-related gratings; (ii) If SHG gratings were present, their QPM bandwidth would be limited to approximately $\delta\lambda_{\rm FWHM}\approx0.44\lambda_{\rm FH}^2/((n_{\rm g} (\lambda_{\rm SH} )-n_{\rm g} (\lambda_{\rm FH})) L_{\rm eff})=0.04$ nm, where $n_{\rm g}$ is the group index, $L_{\rm eff}$ is the effective waveguide length calculated by $L_{\rm eff}=c\tau_{\rm p}/n_{\rm eff}=cQ_{\rm L}/n_{\rm eff} \omega_0=157$ mm. The observed SFG signal, however, is separated by about 5 nm from the SH central wavelength, far beyond the range of SH gratings.  

In addition to realizing solely SFG within a single resonance scan, it is also possible to alternate between SHG and SFG by controlling the temperature to meet both doubly and triply resonant conditions. The SH/SF wavelength span under the fixed pump wavelength can reach up to 11 nm. 

\begin{figure*}[h!]
    \centering
    \includegraphics[width=0.85\textwidth]{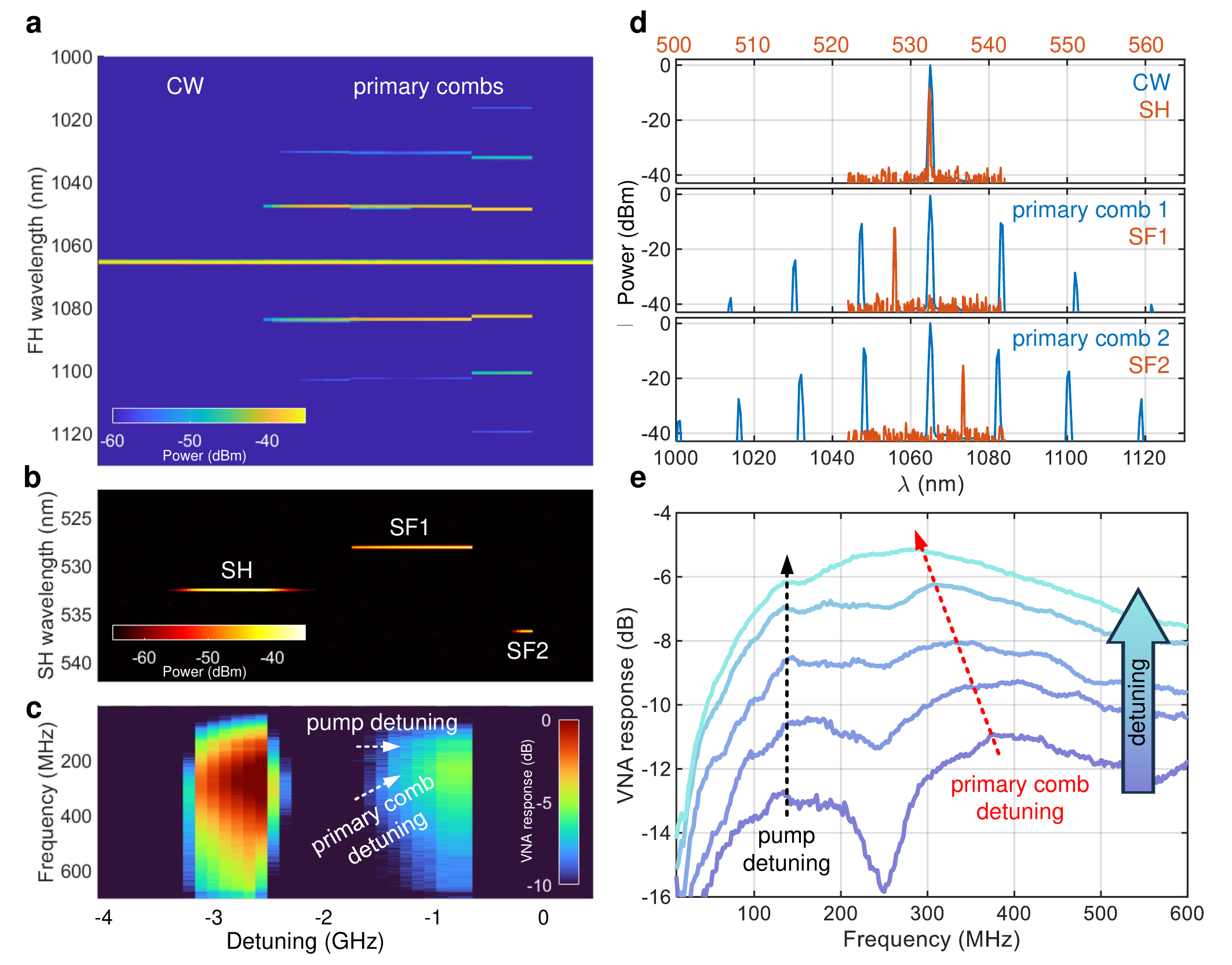}
    \vspace{-5 mm}
    \caption{
    \textbf{Realization and detection of self-assisted SFG.} 
    \textbf{a-c} Evolution of FH, SH/SF, and VNA response in a resonance sweep. FH experiences a CW stage and two primary comb stages. SH is first detected and then replaced by SF in primary comb stages. The VNA response shows single peak at CW stage and shows a dual-peak characteristic in the primary comb stage. 
    \textbf{d} Overlayed dual-axis spectra of FH and corresponding SH/SF. Two self-assisted SF signals with the wavelength difference of 11 nm are realized.  
    \textbf{e} Fine VNA sweep of the SFG generation. The left peak corresponds to FH detuning and the right one reflects the sideband detuning.
    }
\vspace{-2 mm}
    \label{fig:figS7}
\end{figure*}

Figure \ref{fig:figS7} illustrates the transition of second-order frequency transition signals from SH to SF1 and SF2. Figures \ref{fig:figS7}a and \ref{fig:figS7}b depict the spectra of FH and SH/SF, where the SH signal gradually intensifies in the CW state but diminishes as the primary comb emerges. Correspondingly, when the triply resonant condition is met, SF1 signal rapidly occurs under strong coherent pump and disappears promptly after the primary comb transitions. The newly generated primary comb subsequently rewrites the grating, leading to the formation of the SF2 signal. Figure \ref{fig:figS7}d shows the corresponding spectra of FH and SH/SF at three different stages. 

Figure \ref{fig:figS7}c illustrates the evolution of the VNA response, which exclusively detects SH/SF signals. Consequently, no response can be observed in regions where SH or SF are not generated. When SH is generated, the VNA detects a single peak that gradually shifts toward zero frequency (zero detuning). This peak corresponds to the FH detuning, indicating that FH approaches the effective resonance frequency. Due to dominant intrinsic absorption at visable band, the resonance peak of SH/SF cannot be detected.

In contrast to the single-peak SH case, when SF is generated, two resonance peaks are observed, marked by the dashed lines in Figs. \ref{fig:figS7}c and \ref{fig:figS7}e. Figure \ref{fig:figS7}e provides a fine VNA measurement during SFG generation. The position of the left resonance peak remains nearly constant, corresponding to the thermally locked FH signal. The right resonance peak gradually shifts toward zero detuning, corresponding to the sideband of the primary comb. Due to limitations in the VNA scanning speed, the SF2 signal is not detected.

\bibliographystyle{apsrev4-1}
\bibliography{references}